# Did Simon Marius observe Jupiter's satellites on January 8, 1610? An exercise in computation

Yaakov Zik (University of Haifa), Giora Hon (University of Haifa), and Ilan Manulis (Weizmann Institute of Science)

**Abstract**

The question, Did Simon Marius (1573–1625) observe Jupiter's satellites on January 8, 1610 (December 29, 1609 in the Julian calendar) is moot, for he did not disclose his research method and the instrument he used. To resolve this issue we apply astronomical codes and evaluate the visual performance of a replica of the telescope that Galileo Galilei (1564–1642) had used.

**1. Introduction**
The question what Simon Marius (1573–1625) saw when he observed Jupiter's satellites on January 8, 1610 is moot, for he did not disclose his research method and the instrument he used. We apply a methodology which depends on the application of astronomical codes and the evaluation of the visual performance of a replica of the telescope that Galileo Galilei (1564–1642) had used.

We begin with a brief account of the scene of investigation (sect. 2). We continue by sorting out the relevant facts (sect. 3). We then present the astronomical tools we apply and the results of our methodological approach (sect. 4). Finally, we discuss the solution we offer and determine our position vis-à-vis the leading question. We append the paper with the relevant calculations making them available to the interested reader.

**2. The scene of investigation**
Simon Marius (1573–1625) claimed to have observed Jupiter's satellites on December 29, 1609. Marius recorded the date according to the Julian calendar (henceforth dates given in the Julian calendar will be followed by Old Style). The respective date according to the Gregorian calendar, which Galileo Galilei (1564–1642) used when he claimed to have observed Jupiter's satellites, is 10 days later, that is, January 8, 1610. In comparison with Galileo's detailed and systematic observational report in *Sidereus nuncius*, the meagre data provided by Marius makes an in-depth study of his observational report appear to be a futile task. Still, the application of astronomical codes may throw light on the differences between the two sets of data—the one of Marius and that of Galileo. Here is then a claim to be tested, given Marius' spatial and temporal coordinates on January 8, 1610, did he observe in effect Jupiter's satellites?

A recent publication, *Simon Marius and his research*,[1] adds new perspectives on the life and scientific work of Marius. The volume contains a complete English translation of

---
[1] Gaab and Leich (2018).

Marius's *Mundus iovialis anno M.DC.IX detectus ope perspicilli Belgici* (1614).[2] In this book Marius informed the reader that in the summer of 1609 he began observing the heavens with a telescope sent to him from Belgium. On January 8, 1610, while claiming to observe Jupiter, Marius reported that he had seen for the first time three bodies to the West of Jupiter, almost in a straight line with the planet.[3] We recall that Galileo's first observational record of Jupiter's satellites in *Sidereus nuncius* is dated January 7, 1610.[4] Marius's observational report triggered therefore a heated controversy. Galileo was convinced that Marius did not observe Jupiter satellites. Moreover, Galileo thought that Marius' careless observational reports and his lack of understanding of the physical features exhibited by the paths of the satellites around Jupiter, may suggest most probably that Marius never observed them at all.[5]

In a paper titled "Priority, reception, and Rehabilitation of Simon Marius: From the accusation of plagiarism to the Marius-Portal as his virtual collected works," the author, Pierre Leich, claimed that "the history of science has paid little attention to Simon Marius, and he had to wait until the early twentieth century before the quality of his telescopic observations and their independence were finally proved."[6] However, two prominent historians of astronomy, Albert Van Helden and Huib Zuidervaat, cautioned that "the Rehabilitation of Simon Marius" should be assessed carefully. In response to Leich and others, Van Helden and Zuidervaat suggested that "clearly ... further research on Marius's observations is needed before we can begin to speak of a Rehabilitation."[7] Here then is an issue to be decided. Here is where we enter the debate with a view to resolving it and ask first how best to proceed?

**3. Sorting out the facts**
We proceed by exhibiting the uncontroversial historical and physical facts regarding position, time, and the available instruments, which we then presuppose in applying the simulations. We take actual observations with a similar instrument to confirm the feasibility of sighting this phenomenon.

---

[2] Marius (1614): *The World of Jupiter Discovered in the Year 1609 by Means of a Belgian Spy-glass*

[3] Gaab and Leich (2018, pp. 5, 19); Marius (1614, pp. 7r, 18r).

[4] Galileo ([1610] 1989, p. 65); Galileo (1610, pp. 17r–18l). Galileo started recording his lunar observation on November 30, 1609. He recorded the observations of Jupiter from January 7, 1610 until March 2, 1610, less than two weeks before *Sidereus nuncius* was published, see, Drake (1999a, vol. 1, pp. 410–429); Gingerich and Van Helden (2003); Gingerich and Van Helden (2011).

[5] Drake (1960, pp. 165–168); Favaro (1890–1909, vol. 6, pp. 215–217). Marius got a copy of *Sidereus nuncius* in June 1610. The first time Marius's observations were documented in writing was in his *Almanac* of 1611, see, Pasachoff (2018, p. 194); Leich (2018. p. 393); Van Helden and Zuidervaart (2018, p. 415).

[6] Leich (2018, p. 389): The paper covers various aspects of Marius scholarship, e.g., his telescopic observations, arguments for the Tychonic world system, a discussion dedicated to Marius's Rehabilitation, and the foundation of Marius-Portal.

[7] Van Helden and Zuidervaart (2018, p. 415).

*3.1. The appearance of Jupiter*

In general, after sunset the atmosphere is partially illuminated by the sun, being neither totally dark nor completely lit. As a convention, civil twilight begins after sunset and ends when the sun is 6° below the horizon; nautical twilight begins when the sun is 6° below the horizon and ends when the sun is 12° below the horizon; the astronomical twilight begins when the sun is 12° below the horizon and ends when the sun is 18° below the horizon. On January 8, 1610, as darkness fell in Padova and Ansbach, Jupiter was already shining above the horizon crossing over between altitudes of 30° – 50° along the East–Southern sky. Assuming clear sky, cold and steady air, as well as low level of relative humidity at that time,[8] Jupiter must have been bright enough to be seen at or shortly after sunset with the naked eye. The position of the Moon and its illumination may also affect the contrast of an observed object. The Moon at nautical twilight on that day was lagging about 25° to the East side of Jupiter, and its 33 arcminutes disk was 99% illuminated. However, due to the narrow field angle of Galileo's telescope, the contrast at the very surroundings of Jupiter may not have been significantly compromised.

*3.2. Galileo's and Marius's observational reports of January 8, 1610*

Galileo's text in *Sidereus nuncius* was accompanied with figures. On January 8, Galileo followed the same observational sequence he made on January 7, at the first hour of the night,

> On the eighth, I returned to the same observation [which Galileo made in the previous day at the first hour of the night]… I found very different arrangement. For all three little stars were to the West of Jupiter and closer to each other than the previous night, and separated by equal intervals, as shown in the adjoining sketch,[9]

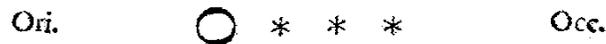

Figure 1

Jean Meeus defined Galileo's first hour of the night as one hour after sunset, that is, at 16:50 UT, about 8 minutes before the end of nautical twilight time in Padova.[10]

Marius did not provide any figure. In his *Mundus iovialis* he wrote,

> My first observation so made was on the 29 December, 1609 [old style]. On that day about 5 o'clock [local time] in the evening I saw three bodies to the West of

---

[8] Cf., fn. 20 below.

[9] Galileo ([1610] 1989, p. 65); Galileo (1610, p. 18l). The time markers Galileo referred to were the time elapsed after the very moment of sunset (*ab occosu*), and the hour of the night (*hora noctis*).

[10] Meeus (1964, p. 105). Note that the time throughout the paper is denoted in UT (Universal Time) units.

> Jupiter, almost in straight line with him. After that I made continues observations till the present time.[11]

Accordingly, the time of Marius's observation in Ansbach was at 16:00 UT, about 16 minutes after sunset (15:44 UT).

It is worth noting that in the early years of the seventeenth century astronomers measured the time from noon when the sun was at the local meridian. Thus, time reckoning during the day was based on the location of the sun and at night on the time elapsed from sunset or the rising of six zodiacal constellations every night which divide the night into six roughly equal parts.[12] However, due to variations in the sizes and positions of the zodiacal constellations and the different seasonal hours which vary in length throughout the year, this rough method could have been trusted to within the time corrected to the nearest hour. But Marius did not report the time in this way. He used a vague term such as *vesperi* which indicates wide time frame stretching (at least) for about 1 hour and 11 minutes (from sunset in Ansbach to the end of the nautical twilight). Marius said nothing about recording time in relation to the location of the sun, sunset, or observation made at night.

### *3.3. Galileo's and Marius's telescopes*

Galileo discussed in his *Sidereus nuncius* the properties of astronomical telescopes.[13] He remarked,

> For it is necessary first that ... [the observers] prepare a most accurate glass that shows objects brightly, distinctly, and not veiled by any obscurity, and second that it multiply ... [the objects] at least four hundred times and show them twenty times closer. For if it is not an instrument such as that, one will try in vain to see all the things observed in the heavens by us and enumerated below.[14]

Galileo singled out four features for an observation to be reliable: 1. the object should be seen bright; 2. the object should be seen distinct; 3. the object should not be veiled by any obscurity; and, 4. the object should be seen at least twenty times closer. In sum, *magnification* and *resolution* are critical for reliable observations.

---

[11] Gaab and Leich (2018, p. 19); Marius (1614, p. 18r): "Interquel illas prima fuit observation 29. Decembris Anni 1609. Quo die vesperi horam circiter quintam tres a Iove occidentales in linea cum ♃ [Jupiter] quasi recta vidi…"

[12] Note that Venus, the evening star, was also used as a time marker and as a useful reference point for measuring the distance between heavenly bodies around early twilight times. However, at the time we are concerned with Venus was about 23.5° below the Western horizon. On methods of telling the time at night, see, Pedersen (1974, 124–132); Evans (1998, 95–99).

[13] On the invention of the telescope, see, Van Helden (1977, pp. 25–28). On Galileo's optical knowledge and telescopes, see, Greco et al (1993); Molesini et al (1996); Ilardi (2007, pp. 207–219); Zik, and Hon (2012); (2014); (2017). On lens-production at the time, see, Bedini (1966); (1967); (1994); Ilardi (2007, pp. 224–235).

[14] Galileo ([1610] 1989, p. 38); Galileo (1610, p. 7l). On the different principles underlying spectacles and telescopes at the time, see, Zik and Hon (2014, 8–12, 19–21).

Marius started his astronomical observations in the summer of 1609. He got a telescope from Belgium with which he observed until January 8, 1610. Some days before January 22, 1610, Marius got two Venetian made glasses, concave and convex, which were fitted into a tube providing an instrument of better qualities than the first instrument he used. But from January 23, 1610, until February 18, 1610, Marius traveled and left the improved instrument at home. Marius reported that he had commenced his astronomical observations of the "Jovian Stars" with the new instrument after February 18, 1610.[15] We do not have any details of Marius' telescopes nor do we know if at all he had the necessary technical skills and optical knowledge for constructing telescopes by himself. Moreover, Marius realized that the local spectacle makers in Ansbach and Nuremberg could not produce lenses suitable for the construction of an improved telescope.[16] He however stopped short of giving any detailed information about the instruments available to him.

It is reasonable then to assume that Marius's first telescope was similar to the one Thomas Harriot (1560–1612) had, that is, a 6 power Galilean telescope. Harriot procured in the Netherlands the telescope with which he made the earliest Moon's observations and drawing in July 26, 1609. Harriot made his first observation of Jupiter's satellites on October 17, 1610, using an improved 20 power Galilean telescope he constructed during the summer of 1610.[17]

Only few Italian observers were able to verify Galileo's discoveries using telescopes made by Galileo and Santini. Notwithstanding Marius's observational reports of early 1610, it stands to reason that not until late September 1610 could he received from Venice superior lenses. According to Van Helden there is no record in the Dutch provinces of any observation of the satellites until 1614.[18] To be sure, many astronomers noticed that even with a rather poor telescope one can see many more stars than with the naked eye (e.g., Harriot, Grienberger, Lembo, Santini, Kepler, Scheiner, etc.). Thus, it stands to reason that Marius was able to see with his first mediocre telescope the planets, the myriad fixed stars, the Milky Way, the Pleiades and so forth.

### 3.4. *What can one see through a telescope*?

We observed Jupiter's satellites using a f/61.2 telescope mounted with replicated lenses of Galileo's 21 power telescope which were manufactured for us by *Elbit Systems, Electro-Optics ELOP Ltd*. Though the lenses were made of modern optical glass they are as close as possible to those of Galileo in terms of their geometrical shape, refractive, and 6dispersive values. The resolution of this telescope is 8.6 arcseconds and its field angle 15 minutes, in comparison to Galileo's f/61.2 telescope which has been tested in 1993 and its estimated resolution was 10 arcseconds, with a field angle of 15 minutes.[19] Our observations which were made in fair-weather conditions,[20] during several successive oppositions of Jupiter as seen in Israel, confirmed the feasibility of Galileo's telescopic observations.[21] Indeed, we were able to observe Jupiter's satellites shortly before the end of the nautical twilight time when the sun was about 11.3° below the horizon in Klil (35° 12' E, 32° 12' N), western Galilee, Israel. In the same manner, it stands to reason that Galileo could have observed Jupiter's satellites on January 8, 1610 shortly before the end of nautical twilight time (16:58 UT) when the sun was about 11.3° below the horizon in Padova.

---

[15] Dick (2018, pp. 159–165); Gaab and Leich (2018, pp. 4–6); Pasachoff (2018, p. 195–197). Note that Marius stayed at Hall (in Swabia) from January 23, 1610 until February 18, 1610.
[16] Ibid.
[17] Shirley (1981, pp. 286–293), (1983, pp. 396–405); North (1981, pp. 241–151).
[18] Van Helden and Zuidervaart (2018, p. 415). Note that telescopes capable of detecting Jupiter's satellites were not available in Europe before September 1610, see Rosen (1965, pp. xvii–xviii); Drake (1999, p. 397); Zik (2004, pp. 486–489); Reeves and Van Helden (2007); Zik and Hon (2014, 16–19).

Needless to say, Mauris' telescopic observations cannot be replicated since no data is available concerning the specifics of the instrument he used.

**4. Results**
The application of astronomical codes creates specific spatiotemporal frames for the occurrences of astronomical events. Such codes provide precise data which can be compared with the sets of the reported observational data, in this case by Galileo and Marius. This comparison yields the following results:

**Galileo**
1. Jupiter was visible to the naked eye at or shortly after sunset (15:50 UT) in Padova (App. 1, below and § 3.4, above).
2. Galileo's figure matches the arrangement of Jupiter's satellites on January 8, 1610 at 17:49 UT (App. 2, below).
3. The beginning of the night was around the end of the nautical twilight time (16:58 UT) in Padova (App. 2, below).
4. Galileo's definition of the first hour of the night was around 17:58 UT (App. 2, below).
5. With f/61.2 Galilean telescope Jupiter's satellites could have been detected around the end of nautical twilight (16:58 UT) when the sun was about 11.3° below the horizon (§ 3.4, above).

**Marius**
1. Jupiter was visible to the naked eye at or shortly after sunset (15:44 UT) in Ansbach (App. 1 below and § 3.4, above).

---

[19] Greco et al (1993, pp. 6222–6223).
[20] We realized that high relative humidity (above 65%) have led directly to increased light scattering, making the sky appear as a white hue, and sometimes even decrease visibility. Such weather conditions also reduced the contrast at the object space, and thus limited the use of telescopes with focal ratio greater than f/35.
[21] Note that Galileo started his observations 30 days after Jupiter's opposition which occurred on December 8, 1609.

2. Ansbach's sky was not dark enough to facilitate the detection of Jupiter's satellites at 16:00 UT (App. 3, below).
3. The telescope Marius used was not good enough to observe Jupiter satellites (§ 3.3, above; App. 3, below).

What do these results mean? The analysis of the spatiotemporal coordinates and the instrumental means available to Galileo and Marius suggest that Marius' claim that he saw Jupiter's satellites on January 8, 1610 could not be the case. In other words, whatever Marius did, whatever he observed, he did not, indeed could not, see these satellites.

## 5. Appendixes

In Appendix 1 we describe the computational tools which we apply, and the observational circumstances in Padova and Ansbach. In Appendixes 2 and 3 we simulate Galileo's and Marius' observations respectively. Two issues stand out:

1. To accurately reconstruct the positions of Jupiter and its satellites on each particular date;
2. To have a better understanding of the actual performance of the telescopes used by Galileo and Marius.

To answer the first issue, we applied JPL's Horizons Web-interface (henceforth, JPL's Horizons), and the astronomical code software package Guide 9.1 (henceforth, Guide). For the second issue, evaluating the actual observational performance of Galileo and Marius, we applied a replica of the 21 power, f/61.2 telescope, which Galileo had used.[22]

Is simulation an accepted methodology for deciding issues in the history of astronomy? Can one rely in this context on a methodology which depends fundamentally on the application of astronomical codes? We are convinced of the strength of this approach. Thus, we present the computed data supporting a plausible solution regarding the question: Given Marius' spatial and temporal coordinates on January 8, 1610, could he have observed Jupiter's satellites? We answer – No.

### Appendix 1

We apply Guide for computing the positions and simulating the appearance of Jupiter's satellites. Guide computes the locations of Jupiter's four largest satellites using the "high accuracy" method described in Jean Meeus' *Astronomical algorithms* 2nd edition of 2009, which, in turn, is based on J. Lieske's E5 theory of the satellites. For computing planetary and lunar positions, we used Guide and JPL's Horizons, which uses DE-431 as a source for generating ephemerides.[23]

---

[22] Drake (1999b, 381, 387).
[23] Guide's accuracy of planetary and lunar positions is good to within a fraction of an arcsecond within the years 1000—3000 AD. As to Jupiter's satellites Guide is indeed good to within better than an arcsecond. See, https://www.projectpluto.com/accuracy.htm.

*Observational circumstances of January 8, 1610*:[24]

|  | **Galileo** | **Marius** |
|---|---|---|
| Location | Padova, Italy | Ansbach, Germany |
| Longitude | 11° 53' 37'' E | 10° 35' 49.2'' E |
| Latitude | 45° 24' 15.3'' N | 49° 19' 48'' N |
| Site | G. Colombo Observatory | --- |
| Altitude ASL | 47m | 405m |
|  |  |  |
| Sunset time | 15:50 UT | 15:44 UT |
| * Sun's azimuth | 238.7° | 236.8° |
| * Sun's altitude | –0.51° | –1° |
|  |  |  |
| End of Civil twilights | 16:21 UT | 16:14 UT |
| End of Nautical twilights | 16:58 UT | 16:55 UT |
| End of Astronomical twilight | 17:34 UT | 17:33 UT |

Table 1: Observational circumstances of January 8, 1610

*Jupiter's positions of January 8, 1610*:

Angular diameter 46 arcseconds; distance from Earth about 4.3 AU.
Jupiter's visual magnitude –2.7. The satellites' visual magnitudes were: Ganymede +4.8, Europa +5.4, Io +5.2, and Callisto +6.

|  | **Padova** | **Ansbach** |
|---|---|---|
| End of Civil twilight | 16:21 UT | 16:14 UT |
| * Jupiter's azimuth | 90.6° | 90.7° |
| * Jupiter's altitude | 32.14° | 30.03° |
|  |  |  |
| End of Nautical twilight | 16:58 UT | 16:55 UT |
| * Jupiter's azimuth | 97.54° | 98.9° |
| * Jupiter's altitude | 38.63° | 36.7° |
|  |  |  |
| End of Astronomical twilight | 17:34 UT | 17:33 UT |
| * Jupiter's azimuth | 105.1° | 107.3° |
| * Jupiter's altitude | 44.83° | 42.73° |

Table 2: Jupiter's positions of January 8, 1610

---

[24] The geographical coordinates were taken from JPL's Horizons. The time throughout the simulations is denoted in UT units. Note that Padova's and Ansbach's time zone is GMT +1 hour thus, for reference, 15:50 UT is equal to 16:50 local time. Note that sunset times are computed according to Guide and JPL's Horizons. However, Guide's sunset and twilight times are rounded to the nearest minute due to the fact that JPL's Horizons does not compute in time intervals less than one minute, while Guide computes the time in fraction of minutes. Whenever the observed object is above the horizon atmospheric refraction is taken into account.

**Appendix 2**

**Galileo** reported that on January 8, 1610, at the 'first hour of the night', all three little stars were to the West of Jupiter, closer to each other and separated by equal intervals, as shown in the adjoining figure,

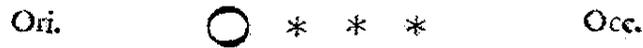

Figure 2

The time of observation at Padova was one hour after sunset, that is, 16:50 UT when the sun was about 10° below the horizon.[25] The simulation below (Fig. 3) presents the locations of Jupiter and the satellites as computed with Guide for 16:50 UT,

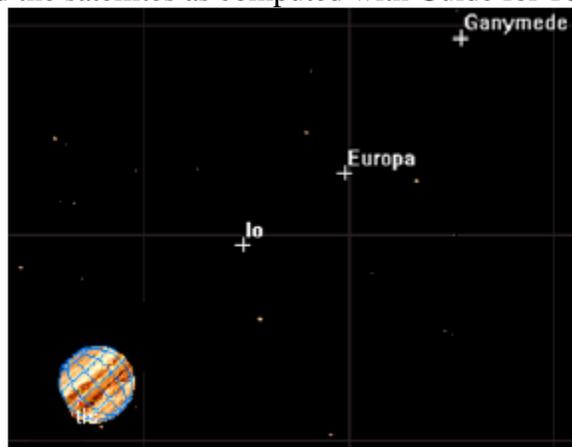

Figure 3

As we can see, the arrangement presented in the simulation of 16:50 UT does not agree with Galileo's figure. Indeed, the satellites are located to the West of the planet but the angular separations between them are not equal. The fourth satellite, located on the East side of the planet, was not seen by Galileo.

The best fit to Galileo's figure is obtained in the simulation made at 17:49 UT (Fig. 4) when the sun was about 20° below the horizon,

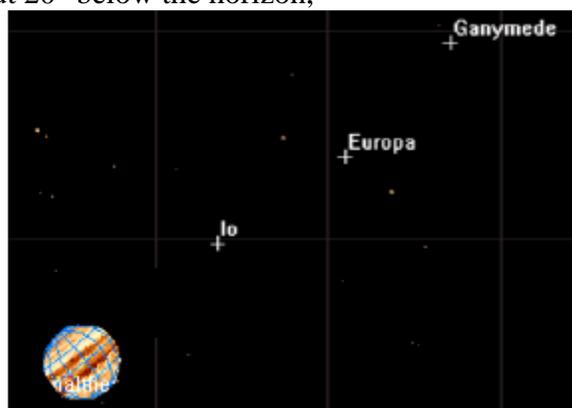

Figure 4

---

[25] Meeus (1964, p. 105).

In this simulation Jupiter's satellites are located to the West of the planet and the fourth satellite, located on the East side, was not seen by Galileo. The computed positions of Jupiter and the satellites in Padova's sky were:
Ganymede    Az. 108.65°;   Alt. 47.41°
Europa      Az. 108.62°;   Alt. 47.4°
Io          Az. 108.6°;    Alt. 47.38°
Jupiter     Az. 108.56°;   Alt. 47.36°
Callisto    Az. 108.37°;   Alt. 47.26° (the satellite does not appear in Galileo's figure).

The angular separation of Jupiter and the satellites, measured between the centers of the heavenly bodies, were:[26]
Ganymede — Europa 88.65" (1.48'); Europa — Io 89.3" (1.49'); Io — Jupiter 106.53" (1.77'); Jupiter — Callisto 604.1" (10.07'), and Callisto — Ganymede 888.58" (14.8').[27]

It is clear that Galileo's reported locations of Jupiter and the satellites could only have been observed at 17:49 UT, about two hours after sunset time in Padova (15:50 UT). This suggests that at the very beginning of his telescopic enterprise Galileo's 'first hour of the night', in effect, was an arbitrary definition of time which did not refer to the actual time when this specific arrangement of Jupiter's satellites was observed and drawn.[28] Accordingly, it appears that Galileo may have considered the beginning of the night about the end of nautical twilight time (16:58 UT) in Padova.

**Appendix 3**

**Marius** reported in *Mundus iovialis* that he observed Jupiter's satellites in Ansbach, on January 8, 1610 about 5 o'clock local time in the evening. Marius's written report did not include any figure but, as mentioned earlier, he saw 3 bodies to the west of Jupiter, aligned almost in a straight line with the planet without specifying any angular separations between

---

[26] Subtracting half the angular radius of Jupiter from the interval between the planet and Io, that is, 106.53"–23"=83.53" (1.4'), makes the apparent angular separation between the two bodies seen as nearly equal to the angular separation between Io – Europa (1.5'), and Europa – Ganymede (1.5').

[27] Note that the 14.8' separation between Callisto and Ganymede is nearly equal to the 15' field angle of Galileo's 21 power telescope. At the time when *Siderius Nuncius* was written, Galileo took the apparent diameter of Jupiter, denoted as one minute, to measure the angular separation between the planet and the satellites. For a discussion on Galileo's diagrams and measurements, see, Drake (1983, pp. 213–223; Zik and Hon (2014, 15–16, 19–21).

[28] Drake and Kowal argued that after January 1612, when Galileo had begun using his new micrometric device for measuring the angular separation between Jupiter and its satellites, his time records could at least be trusted to within 15 minutes, see, Drake and Kowal (1980, 54–55).

them. Marius's observation was made at 16:00 UT, 16 minutes after sunset in Ansbach, when the sun was 3.2° below the horizon. The simulation below (Fig. 5) presents the locations of Jupiter and the satellites as computed with Guide for 16:00 UT,

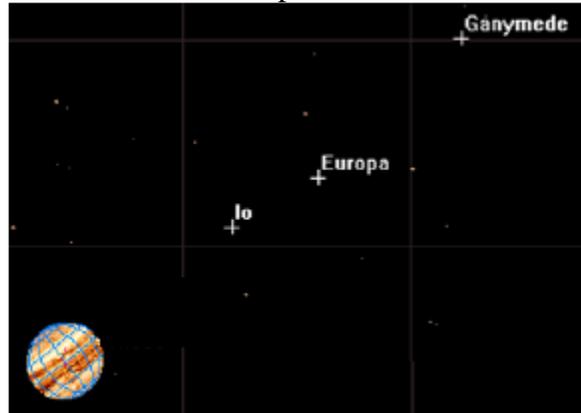
Figure 5

In this simulation the satellites are located to the West of the planet and the fourth satellite, located on the East side, was not reported by Marius. The computed positions of Jupiter and the satellites in Ansbach's sky were:
Ganymede      Az. 88.13°;   Alt. 27.79°
Europa        Az. 88.1°;    Alt. 27.77°
Io            Az. 88.08°;   Alt. 27.76°
Jupiter       Az. 88.05°;   Alt. 27.74°
Callisto      Az. 87.9°;    Alt. 27.64° (Marius did not report seeing this satellite).

The angular separation of Jupiter and the satellites, measured between the centers of the heavenly bodies, were:
Ganymede — Europa 114.74" (1.9'); Europa — Io 57.02" (0.95'); Io — Jupiter 125" (2.08'); Jupiter — Callisto 603.2" (10.05'), and Callisto — Ganymede 899.96" (15').

Due to the different geographical locations of Ansbach and Padova the azimuths and altitudes of Jupiter and the satellites are different. But for the same given time in both places the angular separation between Jupiter and the satellites will always be the same.
   Indeed, Jupiter's three satellites were located on January 8, 1610 to the west of Jupiter, aligned almost in a straight line with the planet. But this arrangement last between 14:45 UT and 20:30 UT. Marius' reported locations of the satellites could only have been observed from 16:55 UT and later, well after 16:00 UT. Recall that at 16:00 UT, about 16 minutes after sunset, Ansbach's sky was not dark enough to make the detection of Jupiter's satellites possible, even with a telescope as good as the 21 power instrument that Galileo used. Marius, who observed the heavens with a telescope since the summer of 1609, must have been aware of the fact that the observational time denoted in his report was not feasible, but he ignored this very fact!

**References**

Bedini, Silvio. 1966. Lens Making for Scientific Instrumentation in the Seventeenth Century. *Applied Optics* 5: 687–694.

Bedini, Silvio. 1967. An Early Optical Lens-Grinding Lathe. *Technology and Culture* 8: 64–80.

Bedini, Silvio. 1994. The Makers of Galileo's Scientific Instruments. II: 10–115. In *Science and Instruments in Seventeenth Century Italy*. Aldershot, UK: Variorum.

Dick, Wolfgang. 2018. Hans Philip Fuch von Bimbach (ca. 1567–1626), Patron of Simon Marius. pp. 139–177. In Gaab, Hans, and Leich, Pierre. Eds. *Simon Marius and His research*. Switzerland: Springer Nature.

Drake, Stillman, and O'mally, C. D., translators. 1960. *The Controversy on the Comets of 1618*. Philadelphia: university of Pennsylvania Press.

Drake, Stillman, and Kowal, Charles. 1980. Galileo's Sighting of Neptune. *Scientific American* 243, 6: 52–59.

Drake, Stillman. 1983. *Telescopes Tides, and Tactics: A Galilean Dialogue about the Starry Messenger and systems of the World.* Chicago: The University of Chicago Press.

Drake, Stillman. 1999. Galileo, Kepler, and Phases of Venus. vol. 1, pp. 396–409. In *Essays on Galileo and the History and Philosophy of Science*. Selected and introduced by Swerdlow N. M., and Levere, T. H. 3 vols. Toronto: University of Toronto Press.

Drake, Stillman. 1999a. Galileo and Satellite Prediction. Vol. 1, pp. 410–429. In *Essays on Galileo and the History and Philosophy of Science*. Selected and introduced by Swerdlow N. M., and Levere, T. H. 3 vols. Toronto: University of Toronto Press.

Drake, Stillman. 1999b. Galileo's First Telescopic Observations. vol. 1, pp. 380–395. In *Essays on Galileo and the History and Philosophy of Science*. Selected and introduced by Swerdlow N. M., and Levere, T. H. 3 vols. Toronto: University of Toronto Press.

Evans, James. 1998. *The History and Practice of Ancient Astronomy*. New York: Oxford University Press.

Favaro, Antonio, ed. 1890–1909, reprinted 1929–39, 1964–66. *Le Opere di Galileo Galilei* [*Opere*], Edizione Nazionale, 20 vols. Florence: G. Barbera.

Galileo, G. (1610). *Sidereus nuncius*. Venetiis: Apud Thomam Baglionum.

Galileo, Galilei. [1610] 1989. *Sidereus nuncius or the Sidereal Messenger* (trans: Van Helden, Albert). Chicago: The University of Chicago Press.

Gingerich, Owen, and Van Helden, Albert. 2003. From Occhiale to printed page: The making of Galileo's *Siderius Nuncius*. *Journal for the history of Astronomy* 34: 251–267.

Gingerich, Owen, and Van Helden, Albert. 2011. How Galileo Constructed the Moons of Jupiter. *Journal for the history of Astronomy* 42: 259–264.

Greco, Vincenzo, et al. 1993. Telescopes of Galileo. *Applied Optics* 32: 6219–6226.

Gaab, Hans, and Leich, Pierre. Eds. 2018. *Simon Marius and His research*. Switzerland: Springer Nature.
Ilardi, Vincent. 2007. *Renaissance Vision from Spectacles to Telescope.* Philadelphia: American Philosophical Society.
Leich, Pierre. 2018. Priority, reception, and Rehabilitation of Simon Marius: From the accusation of plagiarism to the Marius-Portal as his virtual collected works. pp. 389–412. In Gaab, Hans, and Leich, Pierre. Eds. *Simon Marius and His research*. Switzerland: Springer Nature
Meeus, Jean. 1962. Tables of the Satellites of Jupiter. *Journal of the British Astronomical Association* 72, 2: 80–88.
Meeus, Jean. 1964. Galileo's first record of Jupiter's satellites. *Sky & Telescope* February: 105–106.
Molesini, Giuseppe, and Greco, Vincenzo. 1996. Galileo Galilei: Research and Development of the Telescope. pp. 423–438. In Anna, Consortin, ed. *Trends in Optics: Research, Developments and Applications*. St Louis: Academic Press.
North, John. 1981. Thomas Harriot and the First Telescopic Observations of Sunspots. pp. 135–165. In Shirley, John, ed. *A Source Book for the Study of Thomas Harriot*. New York: Arno Press.
Pasachoff, Jay. 2018. Simon Marius *Mundus Iovialis* and the Discovery of the Moons of Jupiter. pp. 191–203. In Gaab, Hans, and Leich, Pierre. Eds. *Simon Marius and His research*. Switzerland: Springer Nature.
Pedersen, Olaf. 1974. *A Survey of the Almagest*. Denmark: Odense University Press.
Reeves, Eileen, and Van Helden, Albert. 2007. Verifying Galileo's Discoveries: Telescope Making at the Collegio Romano. *Acta Historica Astronomia* 33: 1–21.
Rosen, Edward, translator. 1965. *Kepler's Conversation with Galileo's Sidereal Messenger*. New York: Johnson Reprint Corporation.
Shirley, John. 1981. Thomas Harriot Lunar Observations. pp. 284–308. In Shirley, John, ed. *A Source Book for The Study of Thomas Harriot*. New York: Arno Press.
Shirley, John. 1983. *Thomas Harriot: A Biography*. Oxford: Clarendon Press.
Van Helden, Albert. 1977, The Invention of the Telescope. *Transaction of the American Philosophical Society*, vol. 67. Philadelphia: American Philosophical Society.
Van Helden, Albert, and Zuidervaart, Huib. 2018. A word of Caution About the 'Rehabilitation' of Simon Marius. pp. 337–344. In Gaab, Hans, and Leich, Pierre. Eds. *Simon Marius and His research*. Switzerland: Springer Nature
Wright, Ernie. Web Page: *Sidereus nuncius*, Galileo's First Jupiter Observations at: http://www.etwright.org/astro/sidnunj.html.
Zik, Yaakov. 2004. Kepler and the Telescope. *Nuncius*. Firenze: Leo Olschki, XIX: pp. 481–514.
Zik, Yaakov, and Hon, Giora. 2012. Magnification: How to turn a spyglass into an astronomical telescope. *Archive for History of Exact Sciences* 66: 439–464.

Zik, Yaakov, and Hon, Giora. 2014. Galileo knowledge of optics and the functioning of the telescope, revised. https://arxiv.org/ftp/arxiv/papers/1307/1307.4963.pdf.

Zik, Yaakov, and Hon, Giora. 2017. History of Science and Science Combined: Solving a Historical Problem in Optics—the Case of Galileo and his Telescope. Archive for History of Exact Sciences, 4: 337–344.